\documentclass[aps,prd,fleqn,showpacs,showkeys,twocolumn,nofootinbib,
preprintnumbers,superscriptaddress]{revtex4-1}
\usepackage{amsmath,amsfonts,amssymb,amscd,amsxtra,amsthm}
\usepackage{graphicx}  
\usepackage{epstopdf}
\usepackage{color}
\usepackage{dcolumn}  
\usepackage{bm}          
\usepackage{multirow}
\usepackage{slashed}
\usepackage[utf8]{inputenc}
\usepackage[normalem]{ulem} 
\usepackage[dvipsnames]{xcolor} 
\usepackage{array}
\usepackage{slashed}
\renewcommand\sout{\bgroup \color{red} \ULdepth=-.5ex \ULset}

\newcommand\mprdel{\bgroup \color[rgb]{0.8,0.1,0.8} \ULdepth=-.5ex \ULset}

\begin{document}

\preprint{INHA-NTG-02/2017}
\title{Possibility of the existence of charmed exotica}

\author{Hyun-Chul Kim}
\affiliation{Department of Physics, Inha University, Incheon 22212,
  Republic of Korea}
\email{hchkim@inha.ac.kr}

\affiliation{School of Physics, Korea Institute for Advanced Study
  (KIAS), Seoul 02455, Republic of Korea}

\author{Maxim V. Polyakov}
\affiliation{Institut f\"ur Theoretische Physik II, Ruhr-Universit\"at
  Bochum, D--44780, Bochum, Germany} 
\email{maxim.polyakov@tp2.ruhr-uni-bochum.de}

\affiliation{Petersburg Nuclear Physics Institute, Gatchina,
  St. Petersburg 188 300, Russia}

\author{Micha{\l} Prasza{\l}owicz}
\affiliation{M. Smoluchowski Institute of Physics, Jagiellonian
  University, {\L}ojasiewicza 11, 30-348 Krak{\'o}w, Poland}
\email{michal.praszalowicz@uj.edu.pl}

\begin{abstract}
We employ the chiral quark-soliton model to describe excited baryons with one
heavy quark. Identifying known charmed baryons with multiplets allowed by the
model, we argue that apart from regular excitations of the ground state
multiplets, some of recently reported by the LHCb collaboration narrow
$\Omega^{0}_{c}$ states, may correspond to the exotic pentaquarks. This
interpretation can be easily verified experimentally, since exotic $\Omega
^{0}_{c}$ states -- contrary to the regular excitations -- form isospin
triplets, rather than singlets.

\end{abstract}



\maketitle

\section{Introduction}

In a very recent paper the LHCb collaboration announced five, or even six
$\Omega^{0}_{c}$ states with masses in the range of $3 - 3.2$~GeV
\cite{Aaij:2017nav}. Naturally they correspond to the
excitations of the ground state multiplets of charmed baryons that in
this case form two SU(3) sextets: $1/2^{+}$ and $3/2^{+}$. In a recent
paper \cite{Yang:2016qdz} we have shown that these two sextets
together with the ground state $\overline{\mathbf{3}}$ that comprises  
$\Lambda_{c}(2280)$ and $\Xi_{c}(2470)$ can be successfully described in terms
of the chiral quark-soliton model ($\chi$QSM) 
supplemented by an interaction with a heavy quark in such a way
that heavy quark symmetry  \cite{Isgur:1989vq} is respected.
 A great advantage
of the $\chi$QSM consists in a rather restrictive mass formula linking
the spectra of light baryons with the heavy ones in question. 

In the present paper we consider excitations of these ground state multiplets
that are predicted within the $\chi$QSM. They fall into two distinct
categories: the regular excitations that correspond to one-particle
excitation of the initial quark configuration and the exotic ones,
which in the present work are identified with collective rotations
of the soliton~\cite{Diakonov:2010tf, Diakonov:2013qta,
  Petrov:2016vvl}. Since different assignments of the $\Omega^{0}_{c}$
states are possible, we propose criteria that have to be fulfilled by
these excitations. In conclusion we argue that the most probable
assignment is that $\Omega^{0}_{c}(3050)$ and $\Omega 
^{0}_{c}(3119) $ that are very narrow, with the decay widths around 1 MeV,
correspond to the isospin triplet of pentaquarks in the SU(3) $\overline
{\mathbf{15}}$, while the remaining states, including rather wide bumps above 
3.2~GeV, {correspond} to the quark excitations of the ground state
sextets, and are therefore isospin singlets. 

The LHCb discovery triggered several attempts to get an insight into
{the nature of the excited  $\Omega^0_c$'s } in different 
approaches. This includes the QCD sum rules~\cite{Wang:2017zjw,
    Chen:2017sci, Agaev:2017jyt}, the constituent
quark models~\cite{Wang:2017hej}, and lattice QCD
\cite{Padmanath:2017lng}. 
In Refs.~\cite{Karliner:2017kfm,Wang:2017vnc,Cheng:2017ove} 
the new states are treated as bound states of
a charm quark and a light diquark, the authors of
Ref.~\cite{Huang:2017dwn} viewed the new states as 
$\Xi_{c} K$ and $\Xi _{c}^{\prime}K$ molecular states and in some
approaches~\cite{Yang:2017rpg} as pentaquarks. 

The paper is organized as follows. First, we briefly describe the model and
provide formulae for masses {and discuss} the decay widths 
(where possible). Next, we
compare the $\chi$QSM predictions with spectra of excited
$\Lambda_{c}$ and $\Xi_{c}$, and then we discuss possible assignments
of newly discovered $\Omega^{0}_{c}$ states within the pattern of mass
splittings predicted by the model. Finally we conclude, and give
estimates of masses of other members of $\overline{\mathbf{15}}$ and
excited $\mathbf{6}$. 

\section{Chiral Quark-Soliton Model for excited  heavy baryons}
\label{sec:model}

\begin{figure}[h]
\centering
\includegraphics[width=7.9cm]{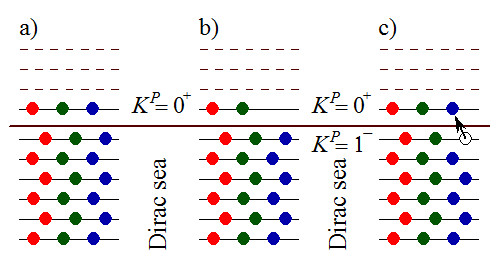} \vspace{-0.2cm}\caption{Schematic
pattern of light ($u$ and $d$) quark levels in a self-consistent soliton
configuration. In the left panel all sea levels are filled and $N_{c}$ (=3 in
the Figure) valence quarks occupy the $K^{P}=0^{+}$ lowest
positive energy level. Unoccupied positive energy levels are dpicted
by dashed lines. In the middle panel one valence quark has been
stripped off, and the soliton has to be 
supplemented by a heavy quark not shown in the Figure. In the right panel a
possible excitation of a sea level quark, conjectured to be $K^{P}=1^{-}$, to
the valence level is shown, and again the soliton has to couple to a heavy
quark. Levels for strange quarks that exhibit different filling pattern are
not shown.}%
\label{fig:levels}%
\end{figure}

The $\chi$QSM is based on an argument of Witten
\cite{Witten:1979kh} that in the limit of large number of colors,
$N_{c}$ relativistic valence quarks generate chiral mean fields
represented by a distortion of a Dirac sea that in turn 
influence the valence quarks themselves (for review see
Ref.\cite{Christov:1995vm}) forming a self-organized configuration called a
\emph{soliton}. Schematic pattern of light quark energy levels corresponding
to this scenario is depicted in Fig.~\ref{fig:levels}.a. Next, rotations of
the soliton, both in flavor and configuration spaces, are quantized
semiclassically and the collective Hamiltonian  is computed. The model
predicts rotational 
baryon spectra that satisfy the following selection rules:

\begin{itemize}
\item allowed SU(3) representations must contain states with hypercharge
$Y^{\prime}=N_{c}/3$,

\item the isospin $\bm{T}^{\prime}$ of the states with $Y^{\prime}%
=N_{c}/3$ couples with the soliton spin $\bm{J}$ to a singlet:
$\bm{T}^{\prime}+\bm{J}=0$.
\end{itemize}

In the case of light parity (+) baryons the lowest allowed representations are
$\mathbf{8}$ of spin 1/2, $\mathbf{10}$ of spin 3/2, and also exotic
$\overline{\mathbf{10}}$ of spin 1/2 with the lightest state corresponding to
the putative $\Theta^{+}(1540)$.

In the recent paper \cite{Yang:2016qdz} following \cite{Diakonov:2010tf} we
have extended this model to baryons involving one heavy quark. In this case
the valence level is occupied by $N_{c}-1$ light quarks (see
Fig~\ref{fig:levels}.b) that couple with a heavy quark $Q$ to form a color
singlet. The first selection rule  in this case reads:
$Y^{\prime}=(N_{c}-1)/3$. Therefore the lowest allowed SU(3) representations
correspond to the soliton of spin 0 in $\overline{\mathbf{3}}$ and spin 1 in
${\mathbf{6}}$. Soliton spin couples with heavy quark spin to form spin 1/2 SU(3)
triplet and two sextets of spin 1/2 and 3/2 that are subject to a hyper-fine
splitting. This pattern is confirmed by the data not only qualitatively but
also quantitatively as shown in Ref.~\cite{Yang:2016qdz}.

The next allowed representation of the rotational excitations
corresponds to the exotic $\overline{\mathbf{15}}$ of spin 0 or spin
1. As we will show below, the spin 1 soliton has a lower mass
and when it couples with a heavy quark it forms spin 1/2 or 3/2 exotic 
multiplets that should be hyper-fine split 
similarly to the ground state sextets. 

The rotational states described above do not change the parity
of the ground state soliton and therefore they correspond to positive
parity. In the present approach negative parity states are generated
by soliton configurations with one light valence quark excited from
the valence level or from the Dirac sea. In this way one can
successfully describe the light baryon spectrum up to 2
GeV~\cite{Petrov:2016vvl}. In this case the second selection rule
above is modified: $\bm{T}^{\prime}+\bm{J}=\bm{K}$, where
$\bm{K}$ denotes so called \emph{grand spin} of the excited valence
quark. Let us remind that the energy levels of the Dirac operator in the
presence of the chiral field with hedgehog symmetry are classified by an
integer $K^{P}$ where $\bm{K}=\bm{l}+\bm{s}%
+\bm{t}$ with $\bm{l}$ standing for quark angular momentum,
$\bm{s}$ for its spin and $\bm{t}$ for isospin~\cite{Christov:1995vm}. $P$ denotes
parity. The soliton configuration with an excited quark develops
its own rotational band.  The
selection rules for excited quark solitons can be therefore summarized as follows:
\begin{itemize}
\item allowed SU(3) representations must contain states with hypercharge
$Y^{\prime}=(N_{c}-1)/3$,

\item the isospin $\bm{T}^{\prime}$ of the states with $Y^{\prime}%
=(N_{c}-1)/3$ couples with the soliton spin $\bm{J}$ as follows:
$\bm{T}^{\prime}+\bm{J}=\bm{K}$, where $\bm{K}$ is the grand spin
of the excited level.
\end{itemize}

The
formula for the soliton mass in the chiral limit for the states in the
SU(3) representation $\mathcal{R}$ has been derived in
Ref.~\cite{Diakonov:2013qta} and reads: 
\begin{widetext}
\begin{equation}
\mathcal{M}^{(K)}   =M^{(K)}_{\text{sol}}+\;\frac{1}{2I_{2}}\left[  C_{2}%
(\mathcal{R})-T^{\prime}(T^{\prime}+1)-\frac{3}{4}Y^{\prime2}\right]
 +\frac{1}{2I_{1}}\left[  (1-a_{K})T^{\prime}(T^{\prime}+1)\frac{{}}{{}%
}+a_{K}J(J+1)-a_{K}(1-a_{K})K(K+1)\;\right] 
 \label{rotmass}
\end{equation}
\end{widetext}
where $C_{2}(\mathcal{R})$ stands for the SU(3) Casimir operator. 
$M_{\text{sol}}^{(K)}\sim N_{c}$ denotes classical soliton mass,
$I_{1,2}$ represent moments of inertia and $a_{K}$
is a parameter that takes into account one-quark
excitation. Although all these parameters can be in principle
calculated in a specific model, we shall follow here a
so called \emph{model-independent} approach 
introduced in the context of the Skyrme model in
Ref.~\cite{Adkins:1984cf}, where all parameters are extracted from the
experimental data.

Note that $a_{K}=0$ if all valence quarks occupy the ground state level and
the soliton spin $J=T^{\prime}$. For solitons constructed from an excited
valence quark configuration $a_{K}\neq0$ and the soliton spin $J$ takes the
following values:
\begin{equation}
J=|T^{\prime}-K|,....,|T^{\prime}+K|.\label{eq:JTprimeK}%
\end{equation}

In the case when the strange quark mass $m_{s}>m_{u,d}\simeq0$,
the soliton mass (\ref{rotmass}) has to be supplemented by the
chiral symmetry breaking Hamiltonian \cite{Diakonov:2013qta}:

\begin{equation}
H_{\mathrm{br}}=\alpha\,D_{88}^{(8)}+\beta\,\hat{Y}+\frac{\gamma}{\sqrt{3}%
}\sum_{i=1}^{3}D_{8i}^{(8)}\,\hat{T}_{i}^{\prime}+\frac{\delta}{\sqrt{3}}%
\sum_{i=1}^{3}D_{8i}^{(8)}\,\hat{K}_{i},\label{eq:Hbr}%
\end{equation}
which has to be evaluated between the collective wave functions~\cite{Diakonov:2013qta,Petrov:2016vvl} 
that depend on
the flavor rotation matrix $A$
\begin{eqnarray}
\Psi_{(\mathcal{R}^{\ast}\,;\,-Y^{\prime}\,T^{\prime}\,T_{3}^{\prime}%
)}^{(\mathcal{R\,};\,Y\,T\,T_{3})}(A)&=&\sqrt{\text{dim}(\mathcal{R}%
)}\,(-)^{T_{3}^{\prime}-Y^{\prime}/2}\notag \\
& \times&D_{(Y,\,T,\,T_{3})(Y^{\prime
},\,T^{\prime},\,-T_{3}^{\prime})}^{(\mathcal{R})\ast}(A)\label{eq:rotwf}%
\end{eqnarray}
coupled to the spin rotational wave function that depends on
the rotational matrix $S$ and to the excited quark function
$\chi_{K_3}$:  
\begin{eqnarray}
&&\Phi_{B,J,J_{3},(T^{\prime},K)}^{(\mathcal{R})}  
 =\sqrt{\frac{2J+1}{2K+1}}%
{\displaystyle\sum\limits_{T_{3}^{\prime},J_{3}^{\prime},K_{3}^{\prime}}}
\left(
\begin{array}
[c]{cc}%
T^{\prime} & J\\
-T_{3}^{\prime} & J_{3}^{\prime}%
\end{array}
\right.  \left\vert
\begin{array}
[c]{c}%
K\\
K_{3}^{\prime}%
\end{array}
\right) \notag \\
&& \times (-)^{-(T^{\prime}+T_{3}^{\prime})}\Psi_{(\mathcal{R}^{\ast
}\,;\,-Y^{\prime}\,T^{\prime}\,T_{3}^{\prime})}^{(\mathcal{R\,};\,B)}%
(A)\,D_{J_{3}^{\prime}J_{3}}^{(J)\ast}(S)\,\chi_{K_{3}^{\prime}},
\label{eq:fullwf}%
\end{eqnarray}
where index $(\mathcal{R\,};\,Y\,T\,T_{3})$
corresponds to the SU(3) quantum numbers of a given baryon in
represntation $\mathcal{R}$, spin index
$(\mathcal{R}^{\ast}\,;\,-Y^{\prime}\,T^{\prime}\,T_{3}^{\prime})$ is
confined to a fixed value of $Y^{\prime}$and formally transforms as a
member of a representation conjugated to $\mathcal{R}$. The
functions $D^{(\mathcal{R})}$ and $D^{(J)}$ are the SU(3) and
SU(2) Wigner matrices, respectively, and
$\chi_{K_{3}^{\prime}}=\left\vert K^{\prime},K_{3}^{\prime}\right\rangle $. 
$\mathcal{O}(m_s)$ parameters $\alpha$, $\beta$, $\gamma$ and $\delta$
are computable in terms of single quark wave functions of valence and
sea quarks. Their explicit form can 
be found in \emph{e.g.} Ref.~\cite{Diakonov:2013qta}.

In order to construct a heavy baryon in the present model we
have to strip off one light quark from the valence level and quantize
the soliton with a new constraint $Y^{\prime}=(N_{c}-1)/3$. The
pertinent light quark configuration is shown in
Fig.~\ref{fig:levels}.b. Such a soliton is coupled with a heavy 
quark to form a color singlet, and the collective Hamiltonian has to be
supplemented by a hyper-fine interaction, which we parametrize as follows~\cite{Yang:2016qdz}:
\begin{equation}
H_{\mathrm{hf}}=\frac{2}{3}\frac{\kappa}{m_{Q}}\bm{J}\cdot
\bm{J}_{Q}\label{eq:ssinter}%
\end{equation}
where $\kappa$ is 
flavor-indepenent. The operators ${\bm{J}}$ and ${\bm{J}}_{Q}$
represent the spin operators for the soliton and the heavy quark,
respectively. 


\section{Phenomenology of heavy baryons\\ in $\chi$QSM}

\label{sect:pheno}

\subsection{Light sector phenomenology}

In order to estimate the heavy baryon masses in the
$\chi$QSM in the model-independent approach one fixes model parameters
from the light sector and uses them for predictions in the heavy quark
sector. This procedure, however, suffers from different systematic
uncertainties. For example, there exist corrections to
$M_{\text{sol}}\sim N_{c}$, that are of the order
$\mathcal{O}(N_{c}^{0})$ related to the Casimir
energy~\cite{Moussallam:1991rj, Scoccola:1998eq} and 
meson loops~\cite{Walliser:1997ee,Nikolov:1996jj, Goeke:2007bj,
    Goeke:2007nc}, which are beyond  
control in the present approach. Obviously, in a model-independent approach
these corrections are accommodated in $M_{\text{sol}}$ and also in $1/I_{1,2}%
$. It is, however, unknown how they depend on the soliton quantum numbers and
how they change in the presence of a heavy quark due to, for example,
nontrivial color interactions between the soliton and an extra quark.

The splittings between multiplets are under much better control
than than the absolute masses. For example, moment of inertia $I_{1}$
can be determined from 
the mass difference of the mean octet ($\mathcal{M}_{\mathbf{8}}
\sim1150$~MeV) and decuplet ($\mathcal{M}_{\mathbf{10}}\sim1380$~MeV)
masses. Indeed, it follows from (\ref{rotmass}):
\begin{equation}
\frac{1}{I_{1}}=\frac{2}{3}(\mathcal{M}_{\mathbf{10}}-\mathcal{M}_{\mathbf{8}%
})=153~\mathrm{MeV},
\end{equation}
which agrees well with much more complete analysis of Ref.~\cite{Yang:2010fm}
giving $1/I_{1}=160$~MeV.

It is, however, much more difficult to estimate the second moment of inertia
$I_{2} $, as it contributes only to the masses of exotic pentaquarks. Given
the fact that the nonexotic members of $\overline{\mathbf{10}}$
can mix with regular baryons~\cite{Goeke:2009ae} $1/I_2$ estimation
suffers from large uncertainty. Also the mass of $\Theta^{+}$, whose
existence is still upheld by the LEPS
\cite{Nakano:2003qx,Nakano:2008ee}, 
the DIANA~\cite{Barmin:2013lva} and a part of 
the CLAS experiment~\cite{Amaryan:2011qc} (see, however, critique in
Ref.~\cite{Anghinolfi:2012yg}), suffers from an uncertainty of 20 MeV: $1520 -
1540$~MeV.  {The  best way to extract  $1/I_2$ is to use the mass of the exotic $\Xi_5$,
since it does not mix with low mass regular hyperons.
Using the values from Refs.~\cite{Goeke:2009ae,Alt:2003vb} we obtain:}
\begin{equation}
\frac{1}{I_{2}}=400 - 450\;\mathrm{MeV}\label{eq:I2}%
\end{equation}
to be compared with even a larger estimate of Ref.~\cite{Yang:2010fm}:
$1/I_{2}=470$~MeV.

Splittings inside SU(3) multiplets are expressed in terms of $\mathcal{O}%
(m_{s})$ parameters: $\alpha$, $\beta$ and $\gamma$. A rather detailed
phenomenological analysis, which includes wave function corrections,
isospin splittings and decay rates, yields rather well constrained
result \cite{Yang:2010fm}, which has been used in
Ref.~\cite{Yang:2016qdz} and which we shall be using here as well:
\begin{equation}
\alpha= -255~\mathrm{MeV}, \;\;\; \beta= -140~\mathrm{MeV}, \;\;\; \gamma=
-101~\mathrm{MeV}. \label{eq:abrNumber}%
\end{equation}

\subsection{Ground state multiplets}

\label{sect:ground}

In order to estimate the masses of the states in
$\overline{\mathbf{3}}$ and $\mathbf{6}$ we have used the
general formula (\ref{rotmass}) with one modification. Since the mean
fields are generated by $N_{c}-1$ valence quarks (see
Fig.~\ref{fig:levels}.b), we have modified $\mathcal{O}(N_{c})$ model 
parameters by the scaling factor $\rho=(N_{c}-1)/N_{c}$, namely: $I_{1,2}
\rightarrow\rho I_{1,2}$ and $\alpha\rightarrow\bar{\alpha}=\rho\alpha$. This
procedure has been applied in \cite{Yang:2016qdz} both for average mass
splittings between the multiplets and for $m_{s}$ splittings within multiplets
of ground state baryons. While the rescaling works very well for $m_{s}$
splittings it is much less accurate for the moments of inertia $I_{1,2}$.
Strictly speaking rescaling by a factor $(N_{c}-1)/N_{c}$ should work well
only for quantities dominated by valence levels, which is probably not the
case for $I_{1}$. Indeed, the rescaling factor that reproduces well
$\mathbf{6}$-$\overline{\mathbf{3}}$ splitting is equal to $\rho=0.9$ rather
than 2/3.

\begin{figure}[h]
\centering
\includegraphics[width=8.4cm]{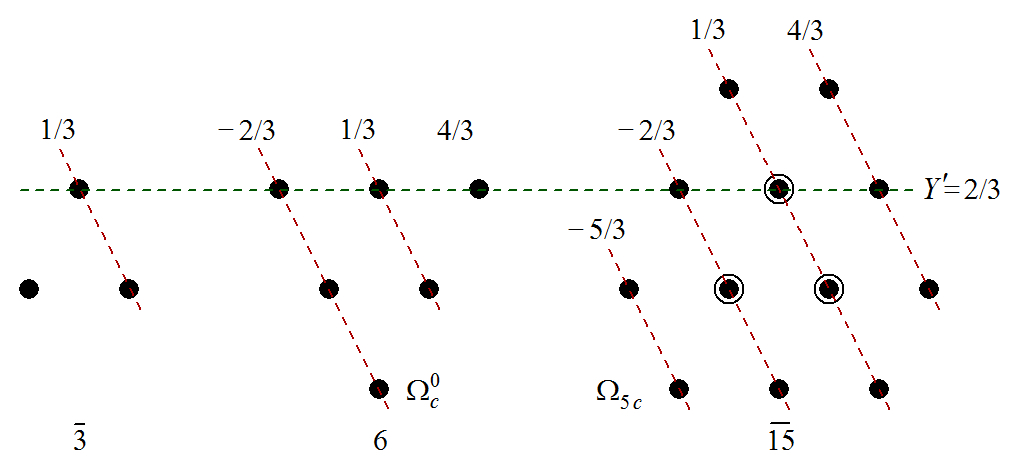} \vspace{-0.2cm}\caption{Rotational
band of a soliton with one valence quark stripped off. The
soliton spin corresponds to the isospin $T^{\prime}$ of states on the
quantization line $Y^{\prime}=2/3$. We show three lowest allowed
representations: the antitriplet of spin 0, the sextet of
spin 1 and the lowest exotic representation $\overline {\mathbf{15}}$
of spin 1 or 0. Diagonal lines indicate the states of equal charges  
(shown above the lines). Heavy quark charge has to be added.}%
\label{fig:reps}%
\end{figure}

Let us briefly summarize the results of Ref.~\cite{Yang:2016qdz}:
\begin{enumerate}
\item Lowest-lying heavy baryons can be indeed grouped in  {two}
  SU(3) multiplets depicted in Fig.~\ref{fig:reps}: an antitriplet of
  spin 1/2 and two sextets of spin 1/2 and 3/2;
\item The sextets are subject to the hyper-fine splitting
  (\ref{eq:ssinter}) that scales like $1/m_{Q}$ and the value of the
  splitting parameter for the charm quark is:
  $\kappa/m_{c} = 70$~MeV; 
\item Within each multiplet $\mathcal{R}$ isospin submultiplets split 
proportionally to the hypercharge: $\delta_{\mathcal{R}} Y$ with parameters
$\delta_{\overline{\mathbf{3}}}= - 180$~MeV and $\delta_{\mathbf{6}}=
-120$~MeV. These values, extracted from the heavy baryon data, are the same
for $b $ and $c$ baryons, they are, however, lower by 11~\% than the values
obtained from the splittings of the light baryon octet and decuplet
with the help of Eq.~(\ref{eq:abrNumber}). This can be 
explained by an 11~\% reduction of the strange quark mass in the
presence of a heavy quark $Q$, since the ratio
$\delta_{\overline{\mathbf{3}}}/\delta_{\mathbf{6}}$ is the same for
both determinations. 

\item Splittings between average $\overline{\mathbf{3}}$ and $\mathbf{6}$
masses are proportional to $1/I_{1}$ and are equal in charm and bottom
sectors. The value of $1/I_{1}$ extracted from heavy baryon spectra and from
the light baryon spectra require tiny rescaling factor $\rho=0.9$.

\item The model predicts a sum rule that links particles from different
multiplets and allows to calculate $\Omega^{*}_{Q}$ mass, which is very well
satisfied for $Q=c$ and gives a prediction for yet unmeasured $\Omega^{*}_{b}$.
\end{enumerate}

\subsection{Exotic $\overline{\mathbf{15}}$ as a rotational excitation}
\label{sec:15bar}

Analogously to the pentaquark $\overline{\mathbf{10}}$ representation, also in
the present case, the soliton admits exotic representations with the
lowest one being $\overline{\mathbf{15}}$ (see
Fig.~\ref{fig:reps}). In this Section we study the properties of heavy 
pentaquarks constructed from a $\overline 
{\mathbf{15}}$ soliton  and a heavy quark. Next possible exotic
representation is $\overline{\mathbf{15}^{\prime}}=(p=0,q=4)$ with
spin $J=1$, which however, is heavier than $\overline {\mathbf{15}}$.

As we can see from Fig.~\ref{fig:reps}, the soliton in $\overline{\mathbf{15}%
}$ can be quantized both as spin $J=0$ and $1$ (remember that the isospin of
the states on $Y^{\prime}=2/3$ line corresponds to spin\footnote{From now on
we use numerical values of the quantum numbers corresponding to $N_{c}=3$,
which does not allow for proper $N_{c}$ counting.}).

In order to estimate the masses of the states in
$\overline{\mathbf{15}}$ we shall use the general formula
(\ref{rotmass}) with the rescaled moments of inertia $I_{1,2} 
\rightarrow\rho I_{1,2}$:
\begin{align}
\mathcal{M}_{\overline{\mathbf{15}},J=0} =  &  M_{\mathrm{sol}} +\frac{5}%
{2}\frac{1}{\rho I_{2}}, \cr \mathcal{M}_{\overline{\mathbf{15}},J=1} = &
M_{\mathrm{sol}} +\frac{3}{2}\frac{1}{\rho I_{2}}+\frac{1}{\rho I_{1}%
}.\label{eq:a15mass}%
\end{align}
Interestingly, the mass difference
\begin{equation}
\Delta_{\overline{\mathbf{15}}}=\mathcal{M}_{\overline{\mathbf{15}}%
,J=0}-\mathcal{M}_{\overline{\mathbf{15}},J=1}= \frac{1}{\rho} \left(
\frac{1}{I_{2}} - \frac{1}{I_{1}} \right)
\end{equation}
is \emph{positive}, since both in the model calculations and model-independent
analyses, $I_{1} \sim3 I_{2}$, which means -- counterintuitively --
that the spin-1 soliton is lighter than the spin-0 one. 

In order to estimate the masses of exotic heavy baryons it is useful
to relate the mean $\overline{\mathbf{15}}$ mass to the mean
$\mathbf{6}$ mass: 
\begin{equation}
\mathcal{M}_{\overline{\mathbf{15}},J=1} =\mathcal{M}_{\mathbf{6}}+\frac
{1}{\rho I_{2}}\label{eq:M15barmass}%
\end{equation}
where we have from \cite{Yang:2016qdz} $\mathcal{M}_{\mathbf{6}}=2580$~MeV.
Given rather large uncertainty of $I_{2}$ (\ref{eq:I2}) and of the factor
$\rho=1-0.66$ we get:
\begin{equation}
\mathcal{M}_{\overline{\mathbf{15}},J=1}= 2980 - 3260~\mathrm{MeV}%
.\label{eq:15bar_range}%
\end{equation}

Finally 
we have to calculate matrix elements of the symmetry-breaking
Hamiltonian 
(\ref{eq:Hbr}). The result reads:
\begin{align}
\Delta_s M_{\overline{\mathbf{15}}}&=Y \left(\beta +
 \frac{17}{144}(\alpha-2\gamma)\right) 
  \\ 
&+\left( -\frac{2}{27} + \frac{1}{24}(T(T+1) -\frac{1}{4}Y^2) \right)
  (\alpha-2\gamma). \notag
\label{eq:mass_anti_15}
\end{align}
Note that in this case the $\delta$ term does not contribute. Using
values from Eq.(\ref{eq:abrNumber}) we obtain
$\delta_{\Omega_{c}}=180$~MeV, which should be further reduced by 
11\% giving $\delta_{\Omega_{c}}=160$~MeV. We therefore predict that
$\Omega_{c}$ from $\overline{\mathbf{15}}$ has mass in the range of $3140 -
3370$~MeV before the hyper-fine splitting, which we estimate using $\kappa
/m_{c}=70$~MeV to be $-50$ and $+20$ MeV for spin 1/2 and 3/2 respectively.
Therefore we see that these rough estimates indicate that some of the states
seen by the LHCb might actually be exotic $\Omega_{5\, c}$
pentaquarks. At this point one should remember that these estimates
are subject to the uncertainties due to the $\mathcal{O}(N_{c}^{0})$
corrections discussed above.

The $\chi$QSM  allows to
estimate the decay widths that proceed through the transition of the
light sector associated with the emission of the 
pseudoscalar meson $\varphi$ ($=\pi,\,K,\,\eta$). The heavy quark
remains in the first approximation intact \cite{Isgur:1989vq}, and
acts merely as a spectator. The 
transition operator can be expressed in terms of three coupling
constants and the collective operators: 
\begin{eqnarray}
\mathcal{O}_{\varphi}&=&3\left[  G_{0}D_{\varphi\,i}^{(8)}-G_{1}\,d_{3bc}%
D_{\varphi\,b}^{(8)}\hat{T}_{c}^{\prime}-G_{2}\frac{1}{\sqrt{3}}D_{\varphi
\,8}^{(8)}\hat{T}_{i}^{\prime}\right] \notag \\
&\times& \frac{p_{i}}{M_{1}+M_{2}}.\label{eq:dec-op}%
\end{eqnarray}
In the present case we will have transitions $\overline{\bm{15}}%
_{1}\rightarrow\overline{\bm{3}}_{0}$ (where the lower index refers to
$T^{\prime}=J$) that includes decays of exotic $\Omega_{c}$ measured
by the LHCb, or $\overline{\bm{15}}_{1}\rightarrow\bm{6}_{1}$ that includes
\emph{e.g.} decays to $\Omega_{c}(2535)+\pi$ that have much larger phase
space. Sandwiching operator (\ref{eq:dec-op}) between rotational wave functions
(\ref{eq:rotwf}) one can calculate the effective decay constants%
\begin{align}
\overline{\bm{15}}_{1} &  \rightarrow\overline{\bm{3}}%
_{0}\qquad G_{\overline{\mathbf{3}}}=G_{0}-\frac{1}{2}G_{1},\nonumber\\
\overline{\bm{15}}_{1} &  \rightarrow\bm{6}_{1}\qquad
G_{\mathbf{6}}=G_{0}-\frac{1}{2}G_{1}-G_{2}.
\end{align}
In this normalization the pion-nucleon decay constant ($g_{\pi
  NN}\sim13$) reads 
\[
g_{\pi NN}=\frac{7}{10}\left(  G_{0}+\frac{1}{2}G_{1}+\frac{1}{14}%
G_{2}\right)  .
\]
Interestingly, in the constituent quark limit \cite{Praszalowicz:1995vi,Diakonov:1997mm}
of the $\chi$QSM
\begin{equation}
G_{0}=(N_{c}+2)G,\;G_{1}=4G,\;G_{2}=2G.
\end{equation}
In the present case, however, due to the fact that we have only $N_{c}-1$
occupied valence levels constant $G_{0}$ should be replaced%
\begin{equation}
G_{0}\rightarrow\bar{G}_{0}=(N_{c}+1)G.
\end{equation}
With this replacement $G_{\mathbf{6}}=0$, an effect similar to the
nullification of the $\Theta^{+}$ width in the same limit
\cite{Diakonov:1997mm}. We therefore expect 
exotic $\overline{\bm{15}}_{1}$ pentaquarks to have small widths, even
if $G_{\overline{\mathbf{3}}}\neq0$ in the  constituent quark limit. We have
cheked that for reasonable set of parameters $G_{0,1,2}$ one can
indeed get the total decay width being {of} the order 
of 1 MeV. 

In the present approach we cannot that easily calculate decay widths that
include $D$ mesons. Fortunately the states that we discuss in this paper 
are lying
below the threshold for such decays.

\subsection{Excited $\overline{\mathbf{3}}$ and $\mathbf{6}$
  multiplets as one-quark excitations}

Possible one quark excitations of the soliton depicted in
Fig.~\ref{fig:levels}.b have been discussed by Diakonov in
Ref.~\cite{Diakonov:2010tf}. By comparing possible excitation energies with the
ones in the light sector, he has come to the conclusion that the most
favourable transition that would lead to excited parity $(-)$ heavy baryons
corresponds to the transition from a $K^{P}=1^{-}$ sea level to an unoccupied
$K^{P}=0^{+}$ state (see Fig.~\ref{fig:levels}.c). Such a transition is not
allowed in the light baryon sector. The very existence of a $K^{P}=1^{-}$
level as a sea level of the highest energy is of course a plausible conjecture
that has to be confirmed by model calculations.

The first allowed SU(3) representation for one quark excited soliton is again
$\overline{\mathbf{3}}$ with $T^{\prime}=0$, which -- according to
(\ref{eq:JTprimeK}) for $K=1$ -- is quantized as spin 1. From (\ref{rotmass})
we have%
\begin{equation}
\mathcal{M}_{\overline{\mathbf{3}}}^{\prime}=M_{\text{sol}}^{\prime}+\frac
{1}{2I_{2}}+\frac{a_1^2}{I_{1}}.
\end{equation}
We will treat $\mathcal{M}_{\overline{\mathbf{3}}}^{\prime}$ as a
phenomenological parameter. Next possibility is flavor $\mathbf{6}$ with
$T^{\prime}=1$, which may couple with $K=1$ to $J=0,1$ and $2$. From
(\ref{rotmass}) we have:
\begin{equation}
\mathcal{M}_{\mathbf{6}\,J}^{\prime}=\mathcal{M}_{\overline{\mathbf{3}}%
}^{\prime}+\frac{1-a_{1}}{I_{1}}+\frac{a_{1}}{I_{1}}\times \left\{
\begin{array}
[c]{rcc}%
-1 & \text{for} & J=0\\
0 & \text{for} & J=1\\
2 & \text{for} & J=2
\end{array}
\right.  .
\end{equation}
Both the $\overline{\mathbf{3}}$ and the $\mathbf{6}$ are
subject to the $m_{s}$ splittings proportional to the hypercharge $Y$. 
For $\overline{\mathbf{3}}$
the splitting parameter is given by the same formula as for the ground state
antitriplet, and therefore we know its numerical value \cite{Yang:2016qdz}:
\begin{equation}
\delta_{\overline{\mathbf{3}}}^{\prime}=\frac{3}{8}\bar{\alpha}+\beta
=\delta_{\overline{\mathbf{3}}}=-180~\mathrm{MeV}.\label{eq:delta3bar}%
\end{equation}
In the case of the sextet the splittings depend on the soliton
spin and read: 
\begin{equation}
\delta_{\mathbf{6}\,J}^{\prime} =\delta_{\mathbf{6}}-\frac{3}{20}\delta
\times \left\{
\begin{array}
[c]{rcc}%
2 & \text{for} & J=0\\
1 & \text{for} & J=1\\
-1 & \text{for} & J=2
\end{array}
\right.  ,
\end{equation}
where $\delta_{\mathbf{6}}=-120$~MeV \cite{Yang:2016qdz}
corresponds to the ground state sextet
splitting. Unfortunately, since we do not know the value of a new parameter
$\delta$, we have no handle on the values of different $\delta_{\mathbf{6}%
\,J}^{\prime}$.

Furthermore, according to Eq.~(\ref{eq:ssinter}), we have to include
the hyper-fine splittings with, however, different
chromomagnetic constant $\kappa^{\prime}$. The model predicts two
SU(3) triplets of spin 1/2 and 3/2, two sextets of spin 1/2 and 3/2
and two sextets of spin 3/2 and 5/2 split by: 
\begin{equation}
\Delta^{\mathrm{hf}}_{\overline{\mathbf{3}}}=\Delta^{\mathrm{hf}}%
_{\mathbf{6}\; J=1}  =\frac{\kappa^{\prime}}{m_{c}},\;\;\;
\Delta^{\mathrm{hf}}_{\mathbf{6}\; J=2}  =\frac{5}{3} \frac{\kappa^{\prime}%
}{m_{c}}%
\label{eq:hf}
\end{equation}
and one sextet, presumably the lightest one, corresponding to $J=0$ with no
hyper-fine splitting.

It is relatively easy to check the $\chi$QSM predictions for excited
$\overline{\mathbf{3}}$, since there are rather well measured candidates.
Indeed for $(1/2)^{-}$ we have $\Lambda_{c}(2592)$ and $\Xi_{c}(2790)$ and for
$(3/2)^{-}$ there exist $\Lambda_{c}(2628)$ and $\Xi_{c}(2818)$. From this
assignment we get $\delta^{\prime}_{\overline{\mathbf{3}}}=-198$~MeV and
$-190$~MeV respectively, in relative good agreement with
Eq.(\ref{eq:delta3bar}). Furthermore, we can extract two other parameters:
\begin{align}
\frac{\kappa^{\prime}}{m_{c}} & = \frac{1}{3}(M_{\Lambda_{c}(2628)}+2
M_{\Xi_{c}(2818)}) \label{eq:kappa'} \\
&-\frac{1}{3} (M_{\Lambda_{c}(2252)}+2 M_{\Xi_{c}(2790)}) =
30~\mathrm{MeV}, \notag \\
\mathcal{M}_{\overline{\mathbf{3}}}^{\prime} & = \frac{2}{9}(M_{\Lambda
_{c}(2628)}+2 M_{\Xi_{c}(2818)})  \\
&+\frac{1}{9} (M_{\Lambda_{c}(2252)}+2
M_{\Xi_{c}(2790)}) = 2744~\mathrm{MeV}. \notag
\end{align}
In the next Section we shall discuss phenomenological application of
the $\chi$QSM to the charmed sextet.

\section{Possible interpretations of the LHCb $\Omega_{c}$ states}

\begin{figure}[h]
\centering
\includegraphics[width=5cm]{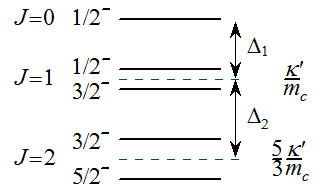} \vspace{-0.2cm}\caption{Schematic
spectrum of excited sextets.}%
\label{fig:spectrum6}%
\end{figure}

The natural scenario, which we will follow in this analysis, is that higher
spin states (or more precisely higher $J$ states) 
become heavier as the spin increases. This assumption leads then to
a $\mathbf{6}$ spectrum depicted 
schematically in Fig.~\ref{fig:spectrum6} with $\left(J=0,\;
  1/2^{-}\right)$ state corresponding to $\Omega_{c}(3000)$. This
spectrum has to be supplemented by two 
possible states $1/2^{+}$ and $3/2^{+}$ belonging to the exotic
$\overline{\mathbf{15}}$. 

The splittings $\Delta_{1,2}$ in Fig.~\ref{fig:spectrum6} correspond
to the $\Omega_{c}$ 
masses for given $J$ before the hyper-fine splitting and read:
\begin{equation}
\Delta_{1}   =\frac{a_{1}}{I_{1}}+\frac{3}{20}\delta,\;\;\;\;
\Delta_{2}   =2\Delta_{1}\label{eq:Deltadiff}.
\end{equation}

The $\chi $QSM predicts five $\Omega_c$ states belonging to the
excited sextet. Therefore we may try to identify {all} five LCHb 
{resonances} with
these states. The corresponding scenario is summarized in
Table~\ref{tab:s1}. We see that in this scenario the relation between
the mass splittings (\ref{eq:Deltadiff}) is badly
broken.This can be further illustrated by observing
that the $\Omega_c$ masses satisfy in the $\chi$QSM two orthogonal sum
rules $\sigma_1=\sigma_2=0$ with: 
\begin{widetext}
\begin{align}
\sigma_1&=6\; \Omega_c(J=0,1/2^-)- \Omega_c(J=1,1/2^-)-8 \;
          \Omega_c(J=1,3/2^-)+3\; \Omega_c(J=2,5/2^-) , \label{sr}\\ 
\sigma_2&=-4\; \Omega_c(J=0,1/2^-)+9\; \Omega_c(J=1,1/2^-)-3 \;
          \Omega_c(J=1,3/2^-)-5 \; \Omega_c(J=2,3/2^-)  
+3\; \Omega_c(J=2,5/2^-),  \notag
\end{align}
\end{widetext}
which are numerically badly violated by the assignment of the minimal
scenario. Let us mention the authors of
Ref.~\cite{Karliner:2017kfm}, who try to interpret the LHCb states
within the quark-diquark model, came to the similar conclusion. 

As can be seen from Table~\ref{tab:s1} the parameter for the
hyper-fine splitting deviates considerably from that determined from
the experimental data for the excited $\overline{\mathbf 3}'$ given in
Eq.~(\ref{eq:kappa'}). Also in this scenario the widths of the would
be hyper-fine  split partners are very different. All these arguments
suggest that such  minimal scenario is not realistic in the mean-field
picture of baryons.  

\renewcommand{\arraystretch}{1.3}
\begin{table}[th]
\begin{center}%
\begin{tabular}
[c]{|c|c|c|c|c|}\hline
$J$ & $S^{P}$ & $M$~[MeV] & $\kappa^{\prime}/m_{c}$~[MeV] & $\Delta_{J}$~[MeV]\\ \hline
0 & $\frac{1}{2}^{-}$ & 3000 & -- & --\\\hline
\multirow{2}{*}{1} & $\frac{1}{2}^{-}$ & 3050 & \multirow{2}{*}{16} &
\multirow{2}{*}{61}\\
~ & $\frac{3}{2}^{-}$ & 3066 &  & \\\cline{1-5}%
\multirow{2}{*}{2} & $\frac{3}{2}^{-}$ & 3090 & \multirow{2}{*}{17} &
\multirow{2}{*}{47}\\
& $\frac{5}{2}^{-}$ & 3119 &  & \\\hline
\end{tabular}
\end{center}
\par
\vspace{-0.3cm} \vspace{0.3cm}\caption{Scenario 1. All LHCb $\Omega_{c}$
states are assigned to the excited sextets. This assignment requires hyperfine
splitting which is almost two times smaller than in the $\overline{\mathbf{3}%
}$ case and relation (\ref{eq:Deltadiff}) is badly broken.}%
\label{tab:s1}%
\end{table}
\renewcommand{\arraystretch}{1}

Given that the minimal scenario does not work, we may try to attribute
some of   the five narrow LHCb $\Omega_c$'s  
to possible exotic $\overline{\mathbf 15}$ multiplet which naturally
emerges in our picture. The states $\Omega_c(3050)$ and
$\Omega_c(3119)$ are good candidates for the $1/2 ^+$ and $3/2
^+$ hyper-fine split $\Omega_c$ members of the $\overline{\mathbf
  15}$. Firstly, the corresponding 
hyper-fine splitting parameter $\kappa/m_c\approx 70$~MeV, in
excellent agreement with the same parameter determined from the
data on the ground-state sextet~\cite{Yang:2016qdz}. Secondly, the widths of
$\Omega_c(3050)$ and $\Omega_c(3119)$ are of order 1~MeV in agreement 
with our expectations discussed in Sect.~\ref{sec:15bar} above. 
The assignment of the LHCb states in this
scenario is summarized in Table~\ref{tab:s2}. 
We see that in this scenario the excited sextet states with
$J=2$ have masses above the $\Xi+D$ threshold at $3185$~MeV,
\textit{i.e.} they can have rather {large} widths
and are not clearly seen in the LHCb data.  

 We have tried several other possibilities to distribute
the observed states over the negative parity
 excited sextet and the positive parity
$\overline{\bm{15}}$, however all of them give less consistent picture.

One can check the suggested identification of new $\Omega_c$ states in
various ways. The simplest {one} would be to search for the {\it isospin}
partners of $\Omega^0_c$ from the $\overline{\mathbf 15}$. For
example, they can be searched in mass distribution of $\Xi_c^0+K^-$  
or $\Xi_c^+ + \bar K^0$, the $\Omega^0_c$'s from the sextet do not decay
into these channels. Another possibility is {to} search for the other
exotic members of the $\overline{\mathbf 15}$ especially the lightest 
$B_c$-baryons (see the next Section). 

\renewcommand{\arraystretch}{1.3}
\begin{table}[th]
\begin{center}%
\begin{tabular}
[c]{|c|c|c|c|c|}\hline
$J$ & $S^{P}$ & $M$~[MeV] & $\kappa^{\prime}/m_{c}$~[MeV] & $\Delta_{J}%
$~[MeV]\\\hline
0 & $\frac{1}{2}^{-}$ & 3000 & -- & --\\\hline
\multirow{2}{*}{1} & $\frac{1}{2}^{-}$ & 3066 & \multirow{2}{*}{24} &
\multirow{2}{*}{82}\\
~ & $\frac{3}{2}^{-}$ & 3090 &  & \\\cline{1-5}%
\multirow{2}{*}{2} & $\frac{3}{2}^{-}$ & \emph{3222} & input & input\\
& $\frac{5}{2}^{-}$ & \emph{3262} & 24 & 164\\\hline
\end{tabular}
\end{center}
\par
\vspace{-0.3cm} \vspace{0.3cm}\caption{Scenario 2. Only three LHCb states are
assigned to {the} sextets. Using relations (\ref{eq:Deltadiff})
and (\ref{eq:hf}) we calculate {the} masses of $J=2$ states (marked in
italics) that fall into a large 
bump seen by the LHCb above 3.2~GeV. In this scenario two narrow states
$\Omega_{c}(3050)$ and $\Omega_{c}(3119)$ are interpreted as exotic
$\overline{\mathbf{15}}$ pentaquarks.}%
\label{tab:s2}%
\end{table}
\renewcommand{\arraystretch}{1}

\section{More on exotic $\overline{\mathbf 15}$}

\subsection{Partners of
  $\Omega_c(3050)$ and $\Omega_c(3119)$ } 

In {the} previous Section we {have} demonstrated that the favourable
scenario is to identify {the} observed narrow {resonances} $\Omega_c(3050)$ and 
$\Omega_c(3119)$ as the $1/2^+$ and $3/2^+$ members of the exotic
$\overline{\mathbf 15}$ multiplet. Now with the help of the mass formula
(\ref{eq:mass_anti_15}) we can predict the masses of other members of
the exotic $\overline{\mathbf 15}$. The parameters of the mass formula
($\alpha,\beta, \gamma$) are fixed by the spectrum of the ground-state
light multiplets (\ref{eq:abrNumber}). {Note that the spectrum has to be
calculated using rescaled $\alpha \rightarrow \overline{\alpha}=2/3\, \alpha$.
Furthermore the splittings have to be reduced by 11\% to account for the
effect discussed in Sect.~\ref{sect:ground}.}
The predicted masses\footnote{Note that the
  predicted masses can be affected by the mixing of non-exotic
  members of the $\overline{\mathbf 15}$ with the ground-state and
  excited $\overline{\mathbf 3}$ and ${\mathbf 6}$, similarily how it
  happens in the light baryon sector, see Ref.~\cite{Goeke:2009ae}.} of
$\overline{\mathbf 15}$ are summarized in
Table~\ref{tab:anti_15}\footnote{We adopt the naming scheme suggested
  by D.I.~Diakonov  \cite{Diakonov:2010tf}}.  {Note that
 with these numbers we get 
$\mathcal{M}_{\overline{\mathbf{15}},J=1}=2935$~MeV,
 just a little below lower limit of Eq.~(\ref{eq:15bar_range}).}

\renewcommand{\arraystretch}{1.3}
\begin{table}[th]
\begin{center}%
\begin{tabular}
[c]{|l|r|c|c|c|}\hline
  &  $Y$&$T$  & $S^{P}=\frac 12^+$ & $S^{P}=\frac 32^+ $\\\hline
$B_c$ & $\frac 53$&$ \frac 12$ & {2685} &2754 \\\hline
$\Sigma_c$ & $\frac 23$&$  1$ & {2808} &{2877} \\\hline
$\Lambda_c$ & $\frac 23$&$  0$ &{2806} &{2875} \\\hline
$\Xi_c$ & $-\frac 13 $&$ \frac 12$ & {2928} &{2997} \\\hline
$\Xi^{3/2}_c$ & $-\frac 13$&$  \frac 32$ & {2931} &{3000} \\\hline
$\Omega_c$ & $-\frac 43$&  {1} & 3050 &3119 \\\hline
\end{tabular}
\end{center}
\par
\vspace{-0.3cm} \vspace{0.3cm}\caption{Predicted masses (in MeV) of
  $1/2^+$ and $3/2^+$ $\overline{\mathbf 15}$-plet
  {under the assumption that}
   $\Omega_c$ members are identified with the observed $\Omega_c(3050)$ and 
  $\Omega_c(3119)$.} 
\label{tab:anti_15}
\end{table}
\renewcommand{\arraystretch}{1}

The exotic $\overline{\mathbf 15}$-plet contains six explicitly exotic
states: $B_c^+, B_c^{++}$ (with {the} minimal quark content $cudd\bar
s$ and $cuud\bar s$), $\Xi_c^{3/2 -},\Xi_c^{3/2 ++}$ ($cdds\bar u$, $cuus\bar
d$), and $\Omega_c^{-},\Omega_c^{+}$ ($cdss\bar u$, $cuss\bar d$).  
The detailed properties of {the} $\overline{\mathbf 15}$-plet we shall study
elsewhere. Here we note that the predicted mass of the lightest
$\overline{\mathbf 15}$ member, the $B_c$-baryon, lies slightly below 
the strong decay threshold to $(\Lambda_c, \Sigma_c)+K$, hence we
predict that the $B_c$-baryon decays only weakly. 

The $\overline{\mathbf 15}$-plet was discussed for the first time by
D.~I.~Diakonov in Ref.~\cite{Diakonov:2010tf}. In this paper the
$\overline{\mathbf 15}$-plet was obtained due to a specific quark
transition between quark levels in the mean-field (an analog of the 
Gamow-Teller transition), so the picture there is different from
ours. In Ref.~\cite{Diakonov:2010tf} the $\overline{\mathbf 15}$-plet
is considerably lighter than in our picture. {For example},
 the mass of the $B_c$ baryon is 2420~MeV. We shall
compare in detail the two pictures elsewhere. 

\subsection{On excited $\Omega_b$}

The mean-field picture of baryons presented here can be easily
generalized to baryons with a bottom quark.  The main feature of our
approach is that the mean field does not depend on the heavy
quark mass.  So,
if we replace the charm quark by the bottom one, we have to make 
an overall shift of the masses and rescale the hyper-fine splittings.   

As for the overall shift of the masses, we take the mass difference of
the ground-state
{antitriplets} for charmed and bottom baryons: 
\begin{equation}
M_{\overline{\mathbf 3}}^b -M_{\overline{\mathbf 3}}^c=3327\,
\mathrm{MeV},  
\end{equation}
which was determined in Ref.~\cite{Yang:2016qdz}, where we 
have also demonstrated that the ratio of the
 hyper-fine mass splittings in the charm and
bottom ground-state sextets is close to $\sim 0.3$, being
in excellent agreement with the mass ratio $m_c/m_b$. 

Performing the overall mass shift and rescaling the hyper-fine
splittings, we obtain the following prediction for the excited
$\Omega_b$: 
$\Omega_b\left({6327},1/2^- \right)$,
$\Omega_b\left(6404,1/2^- \right)$, $\Omega_b\left({6411},3/2^- \right)$,
$\Omega_b\left({6566},3/2^- \right)$ and $\Omega_b\left({6578},5/2^-
\right)$ belonging to {the} excited sextets, and
$\Omega_b\left({6409},1/2^+ \right)$ and $\Omega_b\left({6430},3/2^+
\right)$ belonging to {the} exotic $\overline{\mathbf 15}$-plet. 

\section{Summary and conclusions}

The goal of the  present paper was to classify the $\Omega_c$ baryons
that have been recently reported by the LHCb
collaboration~\cite{Aaij:2017nav}, employing the mean-field approach. 
The mean-field picture of baryons, being justified by the large-$N_c$
limit, offers a unified description of light and heavy baryons. 
We have shown in Ref.~\cite{Yang:2016qdz} that the universal mean field
gives simultaneously good description of the ground-state
$\overline{\mathbf 3}$ and ${\mathbf 6}$ multiplets of heavy 
baryons. Also the ground-state light baryon
multiplets are well described \cite{Petrov:2016vvl}.  In the present
work we have demonstrated that the same picture predicts the following
excited states for heavy-quark baryons in the mass region of
$3000-3200$~MeV:   
\begin{itemize}
\item two hyper-fine split ($1/2^-$ and $3/2^-$)  $\overline{\mathbf 3}'$
which experimentally have very good candidates,
\item five excited sexstets (rotationally and hyper-fine split) with
quantum numbers $(J=0,1/2^-)$, $(J=1,1/2^-,3/2^-)$ and
$(J=2,3/2^-,5/2^-)$, where $J$ denotes the soliton spin, 
\item two hyper-fine split exotic $\overline{\mathbf 15}$-plets
  with quantum numbers $1/2^+$ and $3/2^+$. 
\end{itemize}
Due to the universality of our mean field picture the basic properties
of these excitations are fixed by light baryons  and by ground-state
multiplets of heavy quark baryons. The predictions for the
excited $\overline{\mathbf 3}'$-plets are in excellent agreement with
the experimental spectrum of the excited $\Lambda_c$
and $\Xi_c$.   
 
 The observation of the new excited $\Omega^0_c$'s allows us to get
 insight into the excited sextets and $\overline{\mathbf 15}$-plets. 
We identify the observed $\Omega_c(3000)$, $\Omega_c(3066)$ and
$\Omega_c(3090)$ with $\left(J=0, 1/2^-\right)$ and $\left(J=1,
  1/2^-,3/2^-\right)$ states from the excited sextet, whereas we
identify the most narrow $\Omega_c(3050)$ and $\Omega_c(3119)$ states
with the $\left(J=1, 1/2^+,3/2^+\right)$ states from {the} exotic 
$\overline{\mathbf 15}$ multiplet. The remaining two $\left(J=2,
  3/2^-,5/2^-\right)$ states from the sextet have masses above
$\Xi+D$ threshold (3185~MeV), so they are probably hidden in a large
bump observed by the LHCb collaboration above 3200~MeV. It should be
stressed that the simplest scenario in which all five LHCb $\Omega_c^0$
states are classified as members of the excited sextets, contradicts
general mass formulae derived within the $\chi$QSM.

The simplest way to falsify our identification is to search 
 for the {\it isospin} partners of $\Omega^0_c$ from the $\overline{\mathbf
   15}$. For example, they can be searched in the mass distribution of
 $\Xi_c^0+K^-$ or $\Xi_c^+ + \bar K^0$, the $\Omega^0_c$'s from the
 sextet do not decay into these channels. 

\section*{Acknowledgments}
The authors are grateful to Victor Petrov for illuminating
discussions. H.-Ch. K is grateful to J.Y. Kim and Y.S. Jun for
discussion. The work of H.-Ch.K. was supported by Basic Science
Research Program through the National Research Foundation of Korea
funded by the Ministry of Education, Science and Technology (Grant
Number: NRF-2015R1A2A2A04007048). MVP is thankful to Department of
Theoretical Physics of Irkutsk State University for kind hospitality
and for the support. The work of MVP is supported by CRC110.


\end{document}